# Ultrabroadband Polarization Insensitive Hybrid using Multiplane Light Conversion

**Nicolas K. Fontaine[1,*], Yuanhang Zhang[1,2], Haoshuo Chen[1], Roland Ryf[1], David T. Neilson[1], Guifang Li[2], Mark Cappuzzo[3], Rose Kopf[3], Al Tate[3], Hugo Safar[3], Cristian Bolle[3], Mark Earnshaw[3], and Joel Carpenter[4]**

[1] *Nokia Bell Labs, Crawford Hill Lab, 791 Holmdel-Keyport Rd, Holmdel, NJ, 07733, USA*
[2] *CREOL, the College of Optics and Photonics, University of Central Florida, Orlando, FL 32816, USA*
[3] *Nokia Bell Labs, 600 Mountain Ave., New Providence, NJ 07974, USA*
[4] *The University of Queensland, Brisbane, QLD, 4027, Australia*
*\*E-mail: Nicolas.Fontaine@nokia-bell-labs.com*

**Abstract:** We designed, fabricated and tested an optical hybrid that supports an octave of bandwidth (900-1800 nm) and below 4-dB insertion loss using multiplane light conversion. Measured phase errors are below 3° across a measurement bandwidth of 390 nm. © 2020 The Author(s)

## 1. Introduction

Optical hybrids [1] can be constructed using various technologies including free space, fiber, silicon photonics and using polarization optics. Scaling the usable bandwidth to match the entire range of a photodetector could enable new applications, such as coherent spectroscopy, fiber sensing, LiDAR systems, OCT and other biomedical sensing and imaging systems as well as coherent detection in optical communications. Without active adjustments, the largest reported ultrawide hybrid has a bandwidth of 120 nm around 1550 nm due to difficulties in obtaining an accurate 90-degree phase shift [2-4].

Multiplane light conversion (MPLC) is a multiple input, multiple output beam reshaping technique that consists of multiple phase masks separated by free-space propagation [5, 6] and therefore could produce an optical hybrid which has 2 inputs and 4 outputs. Fig. 1 shows our hybrid MPLC device formed in a multi-bounce cavity comprising 14 phase masks and a gold mirror. The input is fed by a fiber collimator array, and the output are 4 Gaussian beams that are mode matched to a similar fiber collimator array, or which could be detected on a free-space photodetector. Inverse design algorithms [7, 8] are used to optimize each phase mask, and when designed correctly for broadband operation, the phase masks gently perturb the phase structure of the beams which are reshaped after free space propagation. Such devices with very smooth features can be considered adiabatic as the beams slowly change throughout the device. Demonstrations of MPLC are most often used to build mode multiplexers, because the MPLC is a free-space device that has access to the full 2D cross section of the optical fields. However, although rarely demonstrated, it can also be used to emulate most traditional passive devices such as splitters, filters, mode couplers [9], and hybrids [10].

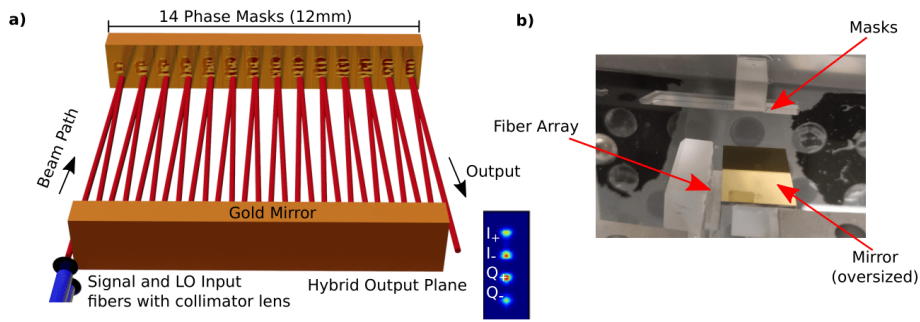

Fig.1 a) Schematic of the 90° optical hybrid using multiplane light conversion. The input beams come from a collimated fiber array. Beams at hybrid output planes are designed to couple into a 250 μm pitch fiber array or a free-space photodetector array using microlenses. The mirror and phase mask are separated by 10 mm. b) Photograph of the assembled MPLC hybrid device.

For building ultra-broadband devices, MPLC has numerous advantages over waveguides and other free-space technologies. Primarily, ultra-broadband operation can be achieved because most of the light just propagates without loss in free space with very gentle phase modifications by reflective phase masks. In particular, it is in principle possible to produce devices operating in the mid infrared or also in the ultraviolet if desired, where waveguides would typically have high losses or require exotic materials. Additionally, the technology to fabricate phase masks is mature and therefore, simulations and designs match almost perfectly with the experimental results. In this paper, we fabricate a device with simulated phase error below 5° across 1 octave of bandwidth between 900 nm and 1800 nm. Measurements were conducted between 1260 nm and 1650 nm due to the limitation of available light sources.

## 2. Design

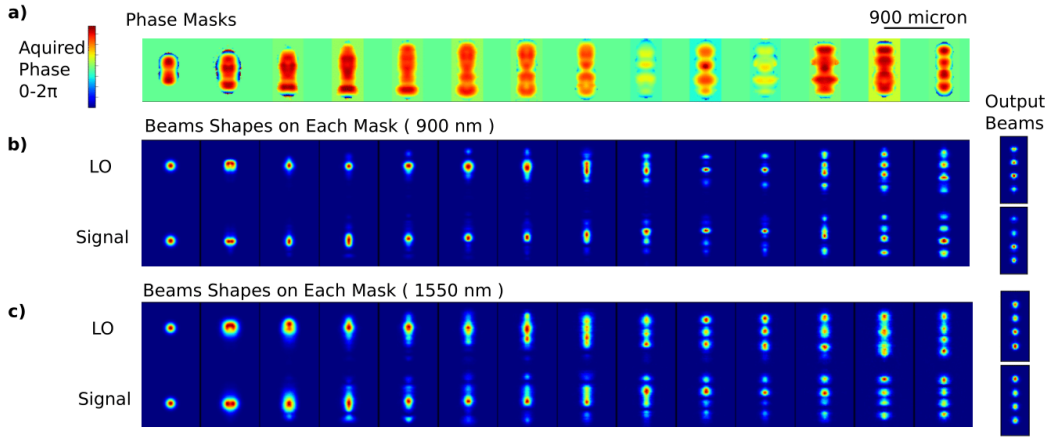

Fig.2 Design and simulation of the optical hybrid at two different wavelengths (900 nm and 1550 nm). a) Phase masks, and beam shapes at each of the planes for b) 900 nm and c) 1550 nm input. Phase masks are relatively flat, lens like, smooth, and have no phase wraps.

Figure 2 shows the design and simulation of the hybrid. An optical hybrid can be described as a spatial multiplexer supporting two spatial modes. The two input modes, local oscillator (LO) input and signal input, are spatially separated Gaussian beams and the two output modes are a composite of 4 equally spaced Gaussian beams with identical power but different phases. In a 90° optical hybrid, the two output modes are orthogonal due to the phase differences between the LO generated beams and signal generated beams that are 0, 180, 90, and 270 degrees. The design process uses inverse design techniques, specifically a modified version of wavefront matching algorithm [8], to optimize modes and corresponding phases at a broad range of wavelengths. The detailed algorithm is described in detail in [11, 12]. The input and output beams are collimated beams from a fiber collimator array with 250 µm pitch, and 70 µm waists placed 15 mm from the first and last planes. The plane spacing is set at 20 mm (10-mm mask-to-mirror distance) and the masks are separated horizontally by 900 µm. Based on simulations, 5 phase masks are sufficient to produce an optical hybrid with 5° phase error in the C-band (50-nm). However, when the number of phase masks is increased, the additional degrees of freedom provided can be used to significantly extend the bandwidth rather than to increase the number of supported modes. Our broadband device uses 14 very smooth phase masks supporting operation between 900 nm and 1800 nm. Each mask imparts much less than π phase across the beams, has no phase wraps and looks similar to a weak distorted lens. It should be noted that a smooth profile alone does not ensure wavelength independence since the phase shift provided by the relief structure changes by a factor of two across an octave of bandwidth. Fig. 2b) and c) show the beam intensity on each phase mask at 900 nm and 1550 nm. The beams slowly transit from a single spot to 4 spatially separated spots at the output. The shorter wavelength tends to be more confined to a spot than the longer wavelength. Because the beams at different wavelengths are distributed differently across the mask, they will experience different phase perturbations at each plane which can partially explain how the bandwidth can be increased.

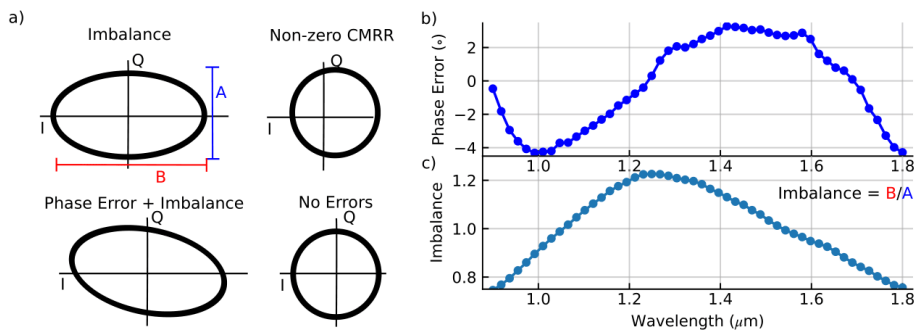

Fig. 3 Simulated performance of the MPLC hybrid. a) Constellations traced by an optical hybrid when two single frequency lasers are input to the LO and signal ports. Imbalance squishes the circle, phase error skews the circle, and non-zero common mode rejection ratio (CMRR) shifts the center of the circle. Simulated hybrid b) phase error and c) imbalance across one octave of bandwidth (900 nm to 1800 nm).

Figure 3a) shows constellation plots when two constant intensity laser sources are input to the LO and Signal ports. We choose to present the constellations because they show all imperfects a hybrid can have. A perfect hybrid traces a circle, a hybrid with a phase error traces a tilted ellipse, a hybrid with only imbalance (different power in I and Q) traces an ellipse aligned to the I/Q axis, and a hybrid with non-zero common mode rejection ratio will not be centered. To characterize the hybrid we extract the imbalance, and the phase error from the constellation. Fig. 3b) shows simulation of the hybrid across the full octave showing phase error below 5° and imbalance between 0.8 and 1.2.

## 3. Fabrication and measurement results

The phase mask was fabricated by 6 subsequent steps of binary mask patterning and etching to give $2^6 = 64$ etch depths between 0 and 775 nm. The spatial resolution of the phase mask was 5-µm × 5-µm. A gold mirror coating was deposited on the phase mask. A plane gold mirror was aligned 10-mm away to fold the beam and allow for the multiple bounces on the phase masks. The input came from two fibers in a 250-µm pitch array which were collimated using micro-lenses attached to the array.

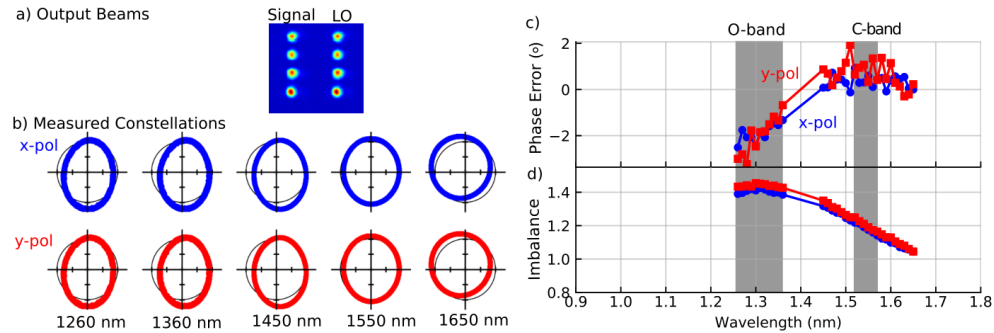

Fig. 4 Measured performance of hybrid between 1260 nm and 1650 nm. a) Output beams when launching light in only the LO or Signal ports, showing spatial profiles with spacing of 250-µm. b) Measured constellations for the two polarizations. Ideal constellation indicated as black line. c) Phase error across the measurement range is below ±3°. Dots indicate measurements. d) Imbalance at the different wavelengths.

The device was measured by launching a laser and a time delayed copy of the laser into the signal and LO port, respectively. Measuring the 390 nm measurement range required two external cavity lasers operating in either the O-band or C+L bands. The output polarizations were split by a Wollaston prism after the MPLC and the 8 spots were measured on an InGaAs camera. For each image, one constellation value for each polarization could be calculated from the intensity of the 8 spots (2 polarizations). 2000 images were taken at each wavelength to acquire enough data to produce a constellation. The phase error was calculated by curve fitting to the measured constellation. Fig. 4a) shows the measured output beams are Gaussian shaped at all the measurement wavelengths. Fig. 4b) shows the constellations at the different wavelengths compared to an ideal constellation (e.g., unit circle). Fig. 4c) and d) plot the curve fitted results for hybrid phase error and imbalance. Towards the shorter wavelengths, there is slight phase error in the hybrid which slightly skews the constellation. The imbalance of this hybrid is much greater than the phase error, however, imbalance is typically fixed by adjusting the gains of the photodetectors whereas phase error cannot be adjusted as simply. Losses are below 4 dB as the dominant loss source is the 2.3 dB of absorption by 27 reflections on the gold mirror.

## 4. Conclusion and Outlook

We designed an optical hybrid with 900 nm (one octave) of 90-degree phase shift and measured performance over 390 nm of less than 3° of phase error. The device is currently 10 mm × 10 mm in size, based on using a 250-µm pitch for input and output which is relatively large compared to integrated devices in high refractive index material such as silicon photonics. Future devices could be much smaller by shrinking the beam sizes [13], moving to a 127-µm pitch array reduces the size to 5-mm×2.5-mm and further reducing the pitch to 50-µm pitch array, would result in a much more compact device of 2.5-mm×0.6-mm.